\documentclass[aps,amsmath,amssymb,superscriptaddress,twocolumn,prl,showpacs,showkeys,array]{revtex4}
\usepackage[dvips]{graphicx}
\usepackage{psfrag}

\newcommand{\be}{\begin{equation}}
\newcommand{\e}{\end{equation}}
\newcommand{\beml}{\begin{subequations}}
\newcommand{\eml}{\end{subequations}}
\newcommand{\beq}{\begin{eqnarray}}
\newcommand{\eq}{\end{eqnarray}}
\newcommand{\ba}{\begin{array}}
\newcommand{\ea}{\end{array}}

\begin{document}

\author{A. E. Miroshnichenko}
\affiliation{Nonlinear Physics Centre, Research School of Physical
Sciences and Engineering, Australian National University, Canberra
ACT 0200, Australia}

\author{M. Schuster}
\affiliation{Physikalisches Institut III, Universit\"at
Erlangen-N\"urnberg, D-91058 Erlangen, Germany}

\author{S. Flach}
\affiliation{Max-Planck-Institut f\"ur Physik komplexer Systeme, N\"othnitzer
Strasse 38, D-01187 Dresden, Germany}

\author{M. V. Fistul}
\affiliation{Theoretische Physik III,
Ruhr-Universit\"at Bochum, 
D-44801 Bochum, Germany}

\author{A. V. Ustinov}
\affiliation{Physikalisches Institut III, Universit\"at
Erlangen-N\"urnberg, D-91058 Erlangen, Germany}

\title{Resonant plasmon scattering by discrete breathers in 
Josephson junction ladders}
\date{\today}

\begin{abstract}
We study the resonant scattering of plasmons (linear waves) by discrete
breather excitations in Josephson junction ladders.
We predict the existence
of Fano resonances, and find them by
computing the resonant vanishing of the transmission coefficient.
We propose an experimental setup of detecting these
resonances, and conduct numerical simulations
which demonstrate the possibility to
observe Fano resonances
in the plasmon scattering by
discrete breathers  in Josephson junction ladders.
\end{abstract}

\pacs{05.45.-a; 74.81.Fa; 42.25.Bs}

\maketitle

\section{Introduction}

Discrete nonlinear Hamiltonian systems generically allow for
spatially localized and time periodic states, {\it discrete
breathers} (DBs), which exist thanks to the interplay between nonlinearity 
and discreteness
\cite{reviews}. 
DBs have been detected and studied experimentally
in interacting Josephson junction systems
\cite{binder00a},
coupled nonlinear optical waveguides \cite{eisenberg98}, 
lattice vibrations in
crystals \cite{swanson99}, antiferromagnetic structures \cite{schwarz99},
micromechanical cantilever arrays \cite{sato03}, Bose-Einstein
condensates loaded on optical lattices \cite{BEC}, layered high-$T_c$
superconductors \cite{hightc}. DBs are predicted also 
to exist in the dynamics of dusty plasma crystals \cite{plasma}.

\begin{figure}[htb]
\vspace{20pt} \psfrag{w}{$\omega_q$}
\psfrag{w-W}{$\omega_q-\Omega$}
\psfrag{w-2W}{$\omega_q-2\Omega$}
\psfrag{w-3W}{$\omega_q-3\Omega$}
\psfrag{w+W}{$\omega_q+\Omega$}
\psfrag{w+2W}{$\omega_q+2\Omega$}
\psfrag{w+3W}{$\omega_q+3\Omega$} 
\includegraphics[width=0.5\textwidth]{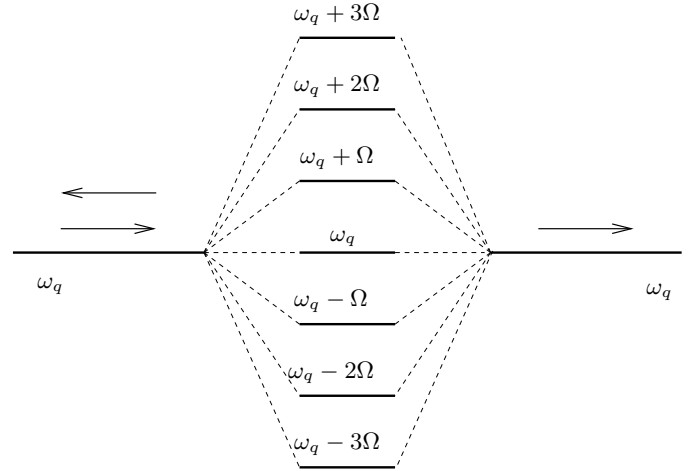}
\caption{Schematic representation of dynamically generated 
paths in the frequency
domain by
a time periodic scattering potential with a period $2\pi/\Omega$. 
In general, there is an infinite number of paths which may lead to  
different interference phenomena.
}
\label{fig3}
\end{figure}

In the transmission
problem of small amplitude waves at frequency $\omega_q$ 
through a DB in the particular
case of a one-dimensional lattice, the DB acts as a time periodic
scattering potential with frequency $\Omega$.
Generally,
there is an infinite number of scattering paths in such a case, which may
lead to a variety of interference phenomena  (see Fig. \ref{fig3}).
Indeed, {\it perfect reflection} was
observed for particular wave numbers
\cite{tcsasf98,kimkim00,kimkim01,sfaemmvf03}. A detailed analysis
of this phenomenon shows \cite{sfaemvfmvf02} that it is a
Fano resonance \cite{fano} which is based on the phenomenon
of destructive interference.

Systems of coupled Josephson junctions with a ladder geometry allow for
the experimental excitation and detection of DBs \cite{binder00a}. 
A variety of experimental
methods allows for a systematic study of DB properties, which have been
successfully combined with theoretical calculations and predictions
\cite{josephsons,etjjmabtpo01,aemsfmvfyzjbp01,aemsfbm03,flach99:_rotob_josep,schuster01:_obser_josep}.
JJLs are quasi-one-dimensional, and allow for the excitation
of travelling linear oscillatory waves (usually referred to
as plasma waves or plasmons), which can be scattered by DBs. 
That makes JJLs  
suitable for the observation of Fano resonances. 
In addition, Josephson junction systems are dissipative systems including both
damping and the presence of
an external homogeneous dc bias. 

The aim of this paper is to investigate the possibility for the observation
of Fano resonances in plasma wave scattering by DBs in Josephson
junction ladders (JJLs). We will also discuss the experimental
setup for the possible observation of these resonances.
We make use of the underdamped regime of dissipation and exploit
previous results on Fano resonances which have been obtained for
nondissipative systems.

\section{Model}

\begin{figure}[htb]
\centering
\includegraphics[width=0.5\textwidth]{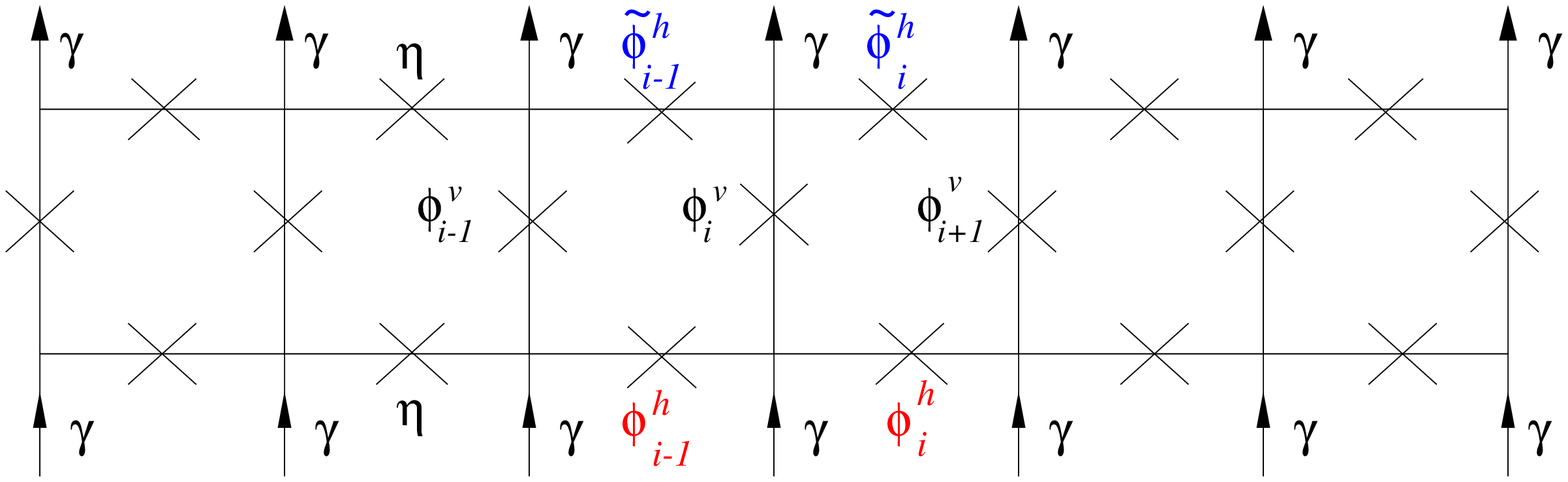}
\caption{Schematic view of the Josephson junction ladder. 
Each cross represents a junction.
Arrows denote the dc bias flow.}
\label{fig4}
\end{figure}

JJLs are formed by an array of small Josephson junctions that are
arranged along the spars and rungs of a ladder, as shown in
Fig.~\ref{fig4}.
%
Junctions are denoted by crosses. Each junction consists of
two small weakly coupled superconducting islands.
The dynamical state of a junction is described
by the phase difference $\phi$ (Josephson phase)
of the superconducting order parameters of the two islands.
When the phase difference does not vary in time
$\phi={\mathrm const}$, the junction is in the superconducting state.
Otherwise, the junction
is in a resistive state with a nonzero voltage drop $V\propto\dot{\phi}$
(the dot represents differentiation with respect to time).

By using the resistively shunted junction (RSJ) model and 
Kirchhoff's laws, we obtain a set of equations for Josephson
junction ladders (JJLs) in the following
form \cite{josephsons,etjjmabtpo01,aemsfmvfyzjbp01}
\begin{eqnarray}\label{jjl}
   \ddot{\phi}_{n}^v+\alpha\dot{\phi}^v_{n}+\sin\phi_{n}^v&
   =&\gamma+\frac{1}{\beta_{L}}(\triangle\phi_{n}^v+\nabla\phi_{n-1}^h
   -\nabla\tilde{\phi}_{n-1}^h)\nonumber \\
   \ddot{\phi}_{n}^h+\alpha\dot{\phi}_{n}^h+\sin\phi_{n}^h &
   =&-\frac{1}{\eta\beta_{L}}(\nabla\phi_{n}^v+\phi_{n}^h-\tilde{\phi}_{n}^h)
   \\
  \ddot{\tilde{\phi}}{}^h_n+\alpha\dot{\tilde{\phi}}{}^h_n+
  \sin\tilde{\phi}_n^h&=&\frac{1}{\eta\beta_{L}}
  (\nabla\phi_{n}^v+\phi^h_n-\tilde{\phi}_{n}^h)\nonumber\;,
\end{eqnarray}
where the notations $\triangle f_n\equiv f_{n-1}-2f_n+f_{n+1}$ and
$\nabla f_n\equiv f_{n+1}-f_n$ are used. The Josephson phases of
vertical, upper and lower horizontal junctions in the $n$th cell
are denoted by  $\phi_n^v$, $\tilde{\phi}{}_n^h$ and $\phi_n^h$
respectively, 
$\alpha$
is an effective damping, $\gamma$ is the external dc bias across vertical
junctions. The discreteness parameter $\beta_L=2\pi I_c^V L/\Phi_0$
characterizes the ratio of geometrical cell inductance $L$ and the Josephson
inductance of vertical junctions, and $\eta=I_c^H/I_c^V$ is the anisotropy
parameter of the Ladder. Here, $I_c^V$ ($I_c^H$) is the vertical
(horizontal) junction critical current (see Ref. 
\cite{schuster01:_obser_josep} for details).

\section{Small amplitude excitations of the superconducting ground state}

First we discuss the
spectrum of small amplitude excitations (plasmons) 
around the superconducting
ground state
$\phi_n^{*v}=\arcsin\gamma$ and
$\phi_n^{*h}=\tilde{\phi}_n^{*h}=0$
by linearizing the system (\ref{jjl})
\begin{eqnarray}\label{lin_gr}
\phi_n^v=\phi^{*v}_n+\varphi_n^v\;,\;
\phi_n^h=\phi_n^{*h}+\varphi_n^h\;,\;
\tilde{\phi}_n^h=\tilde{\phi}_n^{*h}+\tilde{\varphi}_n^h\;,
\end{eqnarray}
where $\varphi_n^v$, $\varphi_n^h$ and $\tilde{\varphi}_n^h$
describe the small amplitude excitations. The presence of
dissipation  leads to the decay of excitations in time. In
experiments, the dissipation $\alpha$ can be rather weak (of
the order of 0.01 or even less) and
usually it is ignored when discussing the small amplitude
excitation spectrum \cite{josephsons}. Here we 
provide the correct solution in the presence of damping. 
We can exclude the
dissipative term by the following time-dependent transformation
\begin{eqnarray} \label{ttrans}
\varphi={\mathrm e}^{-\frac{\alpha}{2}t}\psi\;.
\end{eqnarray}
According to
(\ref{ttrans}) the obtained plane waves are characterized by an
exponential decay in time with a characteristic decay time
$\tau=2/\alpha$.
By solving the equations for $\psi_n \sim e^{i(qn-\omega_q t)}$
we obtain three plasmon bands $\omega_1(q),\omega_2(q),\omega_3(q)$
(see Appendix A). 
One of them
$\omega_1^2(q)=1-\frac{\alpha^2}{4}$ is dispersionless
(i.e. the frequency does not depend on $q$). 
It is characterized by all vertical junctions
being at rest and in-phase excitations of horizontal
junctions in each cell. The  two other plasmon
bands are characterized by
$\omega_2^2(q)<\omega_1^2(q)$ and $\omega_3^2(q)>\omega_1^2(q)$.
Note here, that the band $\omega_2^2(q)$
possesses a rather weak dispersion and becomes dispersionless for $\gamma=0$.
In that limit it also coincides with the first band: 
$\omega_2(q)=\omega_1(q)$ 
for $\gamma=0$.
Therefore, in the Hamiltonian
limit $\alpha=0$ and $\gamma=0$ there is only one plasmon band
$\omega_3^2(q)$ with nonzero dispersion.

\section{Discrete breathers}
\begin{figure}[htb]
\centering
\includegraphics[width=0.35\textwidth]{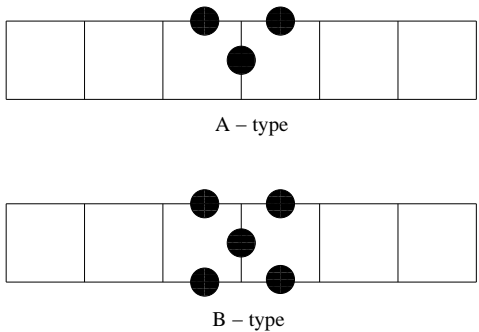}
\caption{Schematic representations of A- and B- type breather resistive cores  
in JJLs with one vertical resistive junction. Black spots mark the
junctions being in the resistive state while all others are
in the superconducting one.}
\label{fig5}
\end{figure}

JJLs support dynamic localized states - discrete
breathers.
A breather is characterized by a few junctions being in the resistive
state $\langle\dot{\phi}\rangle \neq 0$ while the others reside in
the superconducting state $\langle\dot{\phi}\rangle = 0$. The
frequency of a DB is proportional to the average voltage drop
across the resistive junctions
$\Omega_b\propto\langle\dot{\phi}\rangle$ and generally depends on
the parameters of the system. Different
types of DBs have been observed experimentally and numerically
\cite{josephsons}. In the following we focus on 
DB solutions, which are schematically represented in
Fig.~\ref{fig5}.
%

By tuning the external dc bias $\gamma$, a DB state  generally
follows adiabatically and change its characteristics including
voltage drop (frequency), spatial extent etc. We coin such
a state a {\it nonresonant DB}. This is correct unless the DB
solution starts to resonate with the plasmon modes of the
unexcited part of the ladder. Such resonances may either lead to a
loss of the DB solution, or to the appearance of {\it resonant
DBs} \cite{etjjmabtpo01,aemsfmvfyzjbp01,schuster01:_obser_josep}. 
These states are characterized by a
strong coupling of the DB and the extended plasmon modes of the ladder, 
and a
corresponding voltage locking. Consequently resonant DBs are
characterized by their voltage drop (frequency) being nearly
independent of the applied external dc bias. 
Note that resonant DB states do not persist in the dissipationless
Hamiltonian
limit easily, since the absence of dissipation causes a radiation of
the DB energy into the rest of the ladder due to the assumed
resonance with the extended ladder modes. In the dissipative case,
extended modes are damped, and energy is fed into the system
through the breather core, providing with an intricate balance.
However DBs with nondecaying tails in Hamiltonian systems
might well correspond to resonant DBs in dissipative systems.

\section{Dissipationless limit and correspondence idea}

Let us study the Hamiltonian limit, by neglecting both the
dissipation $\alpha$ and the dc bias $\gamma$. In that case we
can later apply a well developed numerical scheme for wave
scattering.
While the wave scattering can be also analyzed for the 
dissipative case, we use the advantages of DB properties
in Hamiltonian systems.
In this limit nonresonant DBs become one-parameter
families of periodic orbits in phase space. Their general form is
\beq\label{db}
\vec{\phi}_n(t)=\vec{\kappa}_n\Omega_bt+\vec{g}_n(t)\;, \eq where
the notation $\vec{f}_n=\{f_n^v,f_n^h,\tilde{f}^h_n\}$ for the
function of cell $n$ is used. The $j$th component of
$\vec{\kappa}_n$ is an integer winding number. It is nonzero for a few
junctions in the DB's core and zero for all others. 
E.g. for the breathers from Fig.~\ref{fig5} we have $\kappa=1$ for all
resistive junctions of the A-breather and zero otherwise, while for
the B-breather $\kappa=1$ for the resistive horizontal junctions and
$\kappa=2$ for the vertical resistive junction.
$\vec{g}_n(t)$
are periodic functions of time
$\vec{g}_n(t~+~\frac{2\pi}{\Omega_b})=\vec{g}_n(t)$ with
exponential decay of amplitudes along the ladder
$\max\limits_t|g^j_{|n|\rightarrow\infty}(t)|~\rightarrow~0$. The
frequency $\Omega_b$ is a parameter which changes along the family of
solutions.

In the presence of weak dissipation $\alpha$ and dc bias
$\gamma$ a given solution from a nonresonant DB family is selected
to become an attractor
and only slightly changed in its temporal evolution.
These changes are small if $\alpha \ll 1$ which we assume.
By tuning the external dc bias $\gamma$
we simply scan dissipative nonresonant DBs which are
close to their nonresonant DBs of the Hamiltonian limit.
For this purpose one can use the relation between $\Omega_b$, $\alpha$
and $\gamma$, which is given in Ref. 
\cite{aemsfmvfyzjbp01}.
This correspondence idea will be exploited in what
follows below.

\section{Scattering of waves by discrete breathers: dissipationless case}

DBs are dynamic localized excitations. For propagating waves
DBs act as time-periodic 
scattering potentials. Based on results for Hamiltonian lattices 
\cite{tcsasf98,kimkim00,kimkim01,sfaemmvf03},
we expect that such interesting phenomena as resonant transmission and resonant
reflection could be observed in JJLs too. 
Here we focus on the resonant reflection, or
Fano resonance.
We start with the Hamiltonian limit, 
where we can rely on recent results on Fano resonances induced by
time periodic scattering potentials \cite{sfaemvfmvf02}. 
We {\it predict} the position
of the resonance and then confirm the prediction using numerical simulations.
The extrapolation of the obtained resonances to the non-Hamiltonian case in the presence of
non zero damping and bias will be shown to be successfull. 
 
In order to study the scattering problem in the Hamiltonian limit
$\alpha=0$ and $\gamma=0$, we linearize the set of
equations (\ref{jjl}) around a DB (\ref{db})
by substituting $\vec{\phi}_n=\hat{\vec{\phi}}_n+\vec{\varphi}_n$
with small amplitude excitations $\vec{\varphi}_n$
\begin{eqnarray}\label{lin}
\ddot{\varphi}_{n}^v+\cos(\hat{\phi}^v_{n})\varphi_{n}^v&=&\frac{1}{\beta_{L}}
(\triangle\varphi_{n}^v+\nabla\varphi_{n-1}^h-\nabla\tilde{\varphi}_{n-1}^h)\nonumber
\\
\ddot{\varphi}_{n}^h+\cos(\hat{\phi}_{n}^h)\varphi_{n}^h&=&-\frac{1}{\eta\beta_{L}}
(\nabla\varphi_{n}^v+\varphi_{n}^h-\tilde{\varphi}_{n}^h)
\\
\ddot{\tilde{\varphi}}{}_{n}^h+\cos(\hat{\tilde{\phi}}{}_{n}^h)\tilde{\varphi}^h_n&=&
\frac{1}{\eta\beta_{L}}(\nabla\varphi_{n}^v+\varphi_{n}^h-\tilde{\varphi}_{n}^h)\nonumber\;.
\end{eqnarray}
This is a set of linear equations with time-periodic
coefficients with the period
$T_b=\frac{2\pi}{\Omega_b}$. 

The set of equations (\ref{lin}) describes scattering of waves by a time-periodic
scattering potential. We may estimate the time-averaged scattering potential
by replacing $\cos(\hat{\phi})=0$ for a resistive junction and  $\cos(\hat{\phi})=1$
for a superconducting junction. 
The interaction part (right hand side of Eq.(\ref{lin})) is not changed.
Due to the finite maximum strength of the scattering potential,
the expected nonresonant scattering
impact can be expected to be weak in general, especially if the plasmon band
width (and thus the corresponding group velocity of waves) is large enough.
That promises ideal grounds for observing resonant reflection on the background
of nearly perfect transmission.

The wave scattering by DBs is studied by using proper boundary
conditions, which implies a scattering setup - at the left hand
side there are incoming and reflected waves and at the right hand
side there are only transmitted ones \beq \vec{\varphi}_n(t)=
\left\{
 \begin{array}{ll}
   \vec{I} {\mathrm e}^{-{\mathrm i}(\omega_3(q) t-qn)}+
   \vec{R}{\mathrm e}^{-{\mathrm i}(\omega_3(q) t+qn)},&n~\ll~0\\
   \vec{T}{\mathrm e}^{-{\mathrm i}(\omega_3(q) t-qn)}~~,& n~\gg~0
 \end{array}
\right. \eq with the frequency $\omega_3(q)$ (\ref{spectr}) and
relation (\ref{polarized}) between the components of the vectors
$\vec{I}$, $\vec{R}$ and $\vec{T}$.

The general solution of (\ref{lin}) can be written as \beq
\vec{\varphi}_n(t) = \sum_{k=-\infty}^{\infty} \vec{B}_{kn}
{\mathrm e}^{-{\mathrm i}(\omega_3(q) + k\Omega_b)t}\;. 
\label{channels} \eq 
Each term
in this sum represents a {\it channel} - the path way for the
waves. There is an infinite number of channels. The $k$th channel
is characterized by its frequency $\omega^k_q=\omega_3(q) +
k\Omega_b$. When the frequency $\omega^k_q$ belongs to the
spectrum (\ref{spectr}) the $k$th channel becomes {\it open},
otherwise it is {\it closed}. It leads to the following property
for the amplitudes $\vec{B}_{kn}$ \beq
\vec{B}_{k,|n|\rightarrow\infty} \left\{ \begin{array}{cl} = 0,&
\textrm{for closed channels}\;.\\
\neq 0,& \textrm{for open channels}
\end{array} \right.
\eq
In other words, plane waves can {\it freely} propagate only
inside the open channels.

The zeroth channel $k=0$ is always open. In principle there is a
possibility to have an additional open channel $\omega^m_q$ with some
nonzero odd integer
$m\not=0$ implying 
\beq \omega_3(q)+\omega_3(q^{\prime})=-m\Omega_b\;.
\eq 
This  multi-channel scattering takes place when
\beq\label{multi}
|\omega_3(0)|<\frac{|m|}{2}\Omega_b<|\omega_3(\pi)|\;. \eq It
corresponds to the case of parametric resonance, which may lead to
the instability of a DB. Moreover, it was shown that multi-channel
scattering is an inelastic process \cite{tcsasf98}. In the
following we restrict our attention only to
the case of one-channel scattering. We can avoid the multi-channel
scattering by choosing frequencies of DBs such that the inequality
(\ref{multi}) is not satisfied.

In the case of one-channel scattering only the zeroth channel
$k=0$ is open and all other channels $k\not=0$ are closed. From
this point of view the scattering potential which is generated by the
DB, can be separated onto two parts: a time averaged dc part (the open channel)
and an ac part (the closed channels). The scattering by the dc part only
is well understood. We are interested in the role of
the set of closed channels which become active inside the breather
core and may provide with interference
phenomena.

Inserting (\ref{channels}) into (\ref{lin}) and eliminating
time leads to a set of equations for the amplitudes $\vec{B}_{kn}$.
These equations yield a characteristic amplitude of the
interaction between the open channel $\vec{B}_{0n}$ and the closed
ones $\vec{B}_{k\neq 0,n}$ inside the DB core, which will be denoted
by $V_{ac-dc}$. If we approximate $\vec{B}_{k\neq 0,n}=0$, 
coefficients in the remaining
equations for $\vec{B}_{0n}$ describe the time-averaged or dc part
of the scattering potential. 

\section{Determining the location of resonances}

As was shown above, a breather acts as a time periodic scattering
potential for propagating waves. The system (\ref{lin}) with time
periodic coefficients with period $T_b$ can be considered as a Floquet problem. 
This system possesses stable solutions, which satisfy
the following condition 
\beq\label{floquet}
\vec{\varphi}_n(t+T_b)={\mathrm e}^{-{\mathrm
i}\theta}\vec{\varphi}_n(t)\;\;. \eq 
These are so-called Floquet
or Bloch eigenstates. Here $\theta$ is a Floquet multiplier, which
can be represented as $\theta=\omega T_b$ with some frequency
$\omega$. Note that, strictly speaking, such frequencies are defined
only modulo $\Omega_b$. Most of all Floquet states are extended ones with 
frequencies $\omega$ belonging to the spectrum $\omega_{1,2,3}(q)$. But due to
the spatial localization of a DB, some Floquet states could be also
localized around a DB with frequencies $\omega$ corresponding
to some localized modes $\omega_L$. 

In order to predict the location of a Fano resonance we use
the results of \cite{sfaemmvf03,sfaemvfmvf02,aemsfbm03}. The 
resonance position is determined by the bound states of the 
dc part of the scattering potential if the width of the
resonance is small compared to the continuum band width. 
That 
condition is equivalent to requesting that
the squared width $1/\beta_L^2$ of 
$\omega_3^2(q)$ is larger than $V^2_{ac-dc}$.
By introducing
the parameter 
\begin{equation}
\lambda_F = \beta_L^2 V^2_{ac-dc} 
\label{weakcoupling}
\end{equation}
it follows $\lambda_F \ll 1$.
In this case the bound states
of the dc part of the scattering potential $\omega_L^{dc}$  
can be considered as
additional discrete levels, which are weakly but resonantly coupled to the open
channel via the frequency of the DB $\Omega_b$. In other words,
the Fano resonance takes place when  
\beq\label{dc}
 \omega_L^{dc}+m\Omega_b=\omega_{q_F}
\eq is satisfied for some $q_F$.
In order to predict the location of the resonance we
have to know the frequencies of the bound states of the 
dc part of the scattering potential. We obtain these values
numerically by diagonalizing the corresponding matrix.

Let us estimate $V_{ac-dc}$ for our case.
In the breather core the $\cos(\Omega_b t)=0.5 (e^{i\Omega_b t}
+ e^{-i\Omega_b t})$ terms in (\ref{lin}) yield
a coupling between channels $k$ and $k\pm 1$ of the order 0.5.
Thus
$V_{ac-dc} \approx 0.5$. Consequently we obtain $\lambda_F =
0.5 \beta_L^2$. Thus for $\beta_L < 1$ we can predict
the position of the Fano resonance as described above. 
Our strategy of searching for Fano
resonances is to compute the frequencies of localized modes of
the dc part of the scattering potential and to satisfy
(\ref{dc}).

\section{Computational results}

For our computations of the transmission coefficient we
studied two types of breathers with {\it left-right} (A-type)
and {\it up-down} (B-type) symmetries with one vertical resistive
junction (see Fig.~\ref{fig5}).

\subsection{The Floquet approach}

We use the formalism described above, and
calculate
the transmission coefficient in the Hamiltonian limit was
by implementing the numerical scheme described in 
Ref.\cite{sfaemmvf03}.
According to that scheme, we 
use proper boundary conditions in the form of a time-periodic
drive with a given frequency from the spectrum (\ref{spectr}). 
We
simulate a system with $N=20$ cells. 
On
the right hand side of the system we use an additional restriction
which implies that there is exactly one transmitted plane wave,
propagating to the right. On the left hand side we do not apply
any additional restrictions. The local time-periodic drive generates
a mixture of waves with different amplitudes
propagating to the left and to the right. In general, we 
excite a superposition of all possible polarization vectors.
However, we drive our system with a frequency from one dispersion
curve. It means, that all waves which correspond to other
dispersion curves decay in space and the desired scattering
setup is realized at some distance from the left boundary. In
our case the two additional branches $\omega_{1,2}(q)$ are
dispersionless and their waves do not propagate through the
system. Therefore, we can already choose the first cell away from
the boundary in order to successfully apply the numerical scheme
from Ref.~\cite{sfaemmvf03}. If we take into account a nonzero damping,
then the seond band $\omega_2(q)$ becomes weakly dispersive as well,
and evanescent modes at these frequencies penetrate slightly
into the ladder. We then simply move our reference sites for
the scattering setup further away from the edges of the ladder,
where again a single frequency excitation is found.

For $\beta_L=0.5$ and $\eta=0.35$ the spectrum (\ref{spectr}) is
located between
$\omega_3(0)=3.53$ to $\omega_3(\pi)=4.52$. We
have found by numerical diagonalization 
that for $\Omega_b=3.1$ all local modes
of
the dc part of the scattering potential
of  A- and B-type breathers with one vertical resistive
junction are located below the plasma frequency $\omega_1$ and there 
are only two of them, 
which satisfy the condition (\ref{dc}):
$\omega_{L1,A}^{dc}=0.62$, $\omega_{L2,A}^{dc}=0.73$,
$\omega_{L1,B}^{dc}=0.93$ and $\omega_{L2,B}^{dc}=0.95$.

\begin{figure}[htb]
\vspace{20pt}
\vspace{20pt}
\includegraphics[angle=270,width=0.5\textwidth]{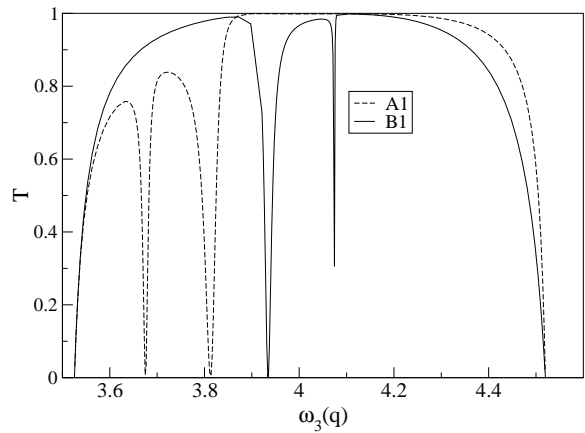}
\caption{
Transmission coefficient versus $\omega_3(q)$ with
Fano resonances for waves scattered by 
A-type (dashed) and B-type (solid) DBs with one vertical resistive junction
for $\Omega_b=3.1$, $\beta_L=0.5$ and $\eta=0.35$.}
\label{fig6}
\end{figure}

Thus we expect to observe two Fano resonances for the A-type
breather at frequencies $\omega_{q_F}=3.72$ and $\omega_{q_F}=3.83$ 
according to (\ref{dc}) and two for the B-type breather 
at frequencies $\omega_{q_F}=4.03$ and $\omega_{q_F}=4.05$.
Indeed, the computation of the
transmission coefficient 
shows that there are two Fano resonances for the
A-type and B-type breathers (Fig.~\ref{fig6}). The frequencies of the
numerically observed resonances 
are $\omega_{q_F}=3.68$ and $\omega_{q_F}=3.81$ for the A-type breather and
$\omega_{q_F}=3.93$ and $\omega_{q_F}=4.07$ for the B-type breather
(see Fig.~\ref{fig6}).
%

The number of localized modes of the dc part of the scattering potential 
does not change when
the damping and bias become nonzero. 
But in this case there are two dispersion curves with nonzero
dispersion $\omega_{2,3}(q)$. One of them is located 
above the plasma frequency $|\omega_1(q)|<|\omega_3(q)|$ 
(as before) and the second one below it $|\omega_2(q)|<|\omega_1(q)|$. 
Since the latter one possesses a rather weak dispersion, 
for small damping
$\alpha\approx0.01$ there are still localized modes, 
which satisfy the condition (\ref{dc}).

We also  
incorporated the finite damping 
into the calculation of the transmission coefficient
by using the transformation (\ref{ttrans}).
The resulting scattering problem can be again analyzed along
the lines of the Floquet approach from above.
Note that this modified scheme implies the 
excitation of a scattering wave setup which decays
in time in a spatial homogeneous way.
The numerical results of this scheme for $\alpha=0.01$ 
coincide with the transmission shown in
Fig.~\ref{fig6}. 
The Fano resonance positions are practically the same, 
because the frequencies of localized modes almost did not change.

\subsection{Direct numerical simulations}
\label{sec:dirnumsim}

%
\begin{figure*}[tb]
  \centering
  \includegraphics[width=1.8\columnwidth]{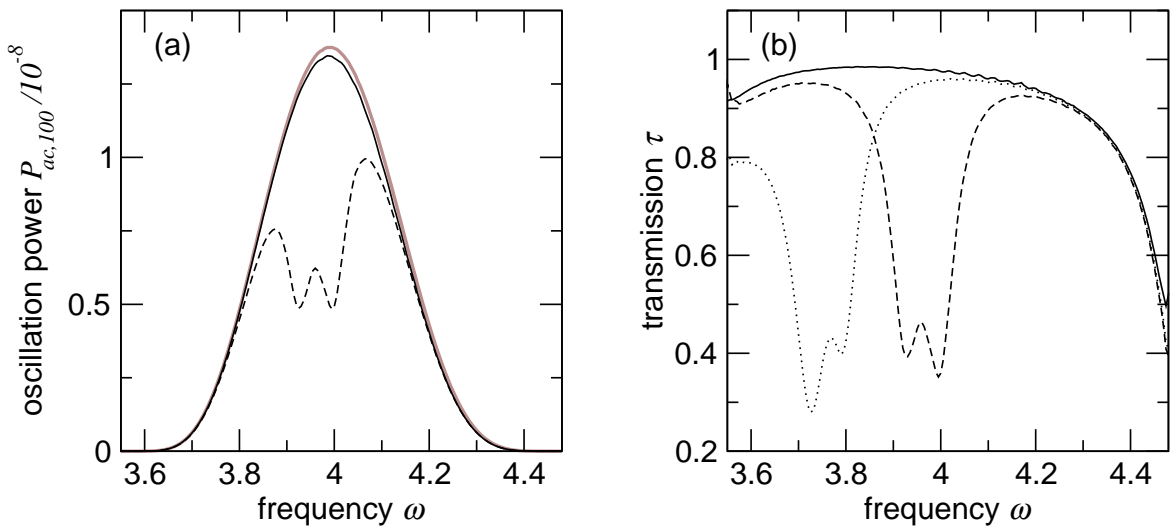}
  \caption{Direct numerical simulation of the linear wave propagation
    in a JJL with $N=100$ cells, $\alpha=0.05$, $\beta_L=0.5$,
    $\eta=0.35$ with a boundary ac bias
    $\gamma_1=\gamma_\text{ac}\cos(\omega t)$. 
    (a) Oscillation power of vertical junction $n=101$ versus
    excitation frequency $\omega$ for an empty system (solid grey line), 
    for an A DB at site $n=50$ with frequency $\Omega_b=2.777$ (solid
    line), and $\Omega_b=3.284$ (dashed line). (b) Transmission
    coefficient $\tau$ for an A DB of frequency 
    $\Omega_b=2.777$ 
    (solid line), $\Omega_b=3.284$
    (dashed line), and frequency $\Omega=3.082$ (dotted line). }.

  \label{fig:rbtransm_marcussimu}
\end{figure*}

In order to test whether the Fano resonances computed 
along the lines of the Floquet approach (see Fig.~\ref{fig6})
also persist in an experimental situation,
we performed direct numerical simulations of the scattering of linear
waves by a DB in Josephson ladders which emulate a real experimental
situation. The equations of motion
were numerically integrated using a standard fifth order Runge Kutta
scheme \cite{press97:_numer} with precision monitoring.
To study the scattering of linear waves by a DB in a situation
comparable to an experiment, a special scattering
experiment scheme was set up. 
Linear waves are generated in a JJL with open ends
by locally applying a 
time-periodic current $\gamma_1(t)$ at the vertical junction $1$:
$
  \gamma_1(t)=\gamma_{ac}\cos(\omega t).
$
The local current acts as a local parametric drive. It excites
a tail of junctions that oscillate with frequency $\omega$. This tail 
extends into the ladder and decays exponentially in space.
The situation is comparable to a DB with frequency
$\omega$ located at the system boundary. In
Refs.~\cite{flach99:_rotob_josep,aemsfmvfyzjbp01}, the spatial extent
of 
oscillatory tails of DBs was computed analytically. If the frequency $\omega$
lies  \emph{inside} the dispersive $\omega_3$ band, the tail 
decays as $\varphi_n\propto\exp(-\alpha n)$. \emph{Outside} the band, the
oscillations practically do not penetrate into the system.
To monitor the linear wave propagation in the system, we
compute the
time-averaged oscillation
power of vertical junction $n$,
$P_{\text{ac},n}=\langle\dot\varphi_n^2\rangle$.

As a scatterer, a DB of either type A or B is launched in the center of
the system. The DB frequency is then tuned by a spatially
uniform DC
bias $\gamma$.

We simulate an open-ended JJL with $N=100$ cells, damping
$\alpha=0.05$, discreteness $\beta_L=0.5$, and anisotropy
$\eta=0.35$. The vertical junction at the left edge $n=1$ is driven by a
time-dependent current with $\gamma_\text{ac}=0.05$. The driving
frequency is swept within $3\leq\omega\leq5$. At each frequency step,
the system is integrated over a time period of
$500$ units to allow for relaxation. The dynamical values are then
monitored over the following $300$ time
units. During that time the oscillation power 
and phase velocities are averaged and recorded.

Fig.~\ref{fig:rbtransm_marcussimu}(a) displays the obtained
average oscillation power spectrum in the vertical junction $n=101$ (at
the right edge of the ladder which is opposite to the ac-driven junction). 
In the empty
case \emph{without} a scattering DB state inside the ladder, the
oscillation power shows a broad peak when the excitation frequency
$\omega$ is inside the $\omega_3$ band. Outside the band, the
oscillations quickly decay to a level below $10^{-20}$. As a test,
the frequency-dependent spatial decay was compared to the analytical
predictions and perfect agreement was found. 
When a scattering DB is inserted into the system at the central site
$n=50$, the propagation of linear waves is modified, and dips may appear
in the power spectrum (dashed line in Fig.~\ref{fig:rbtransm_marcussimu}.

We obtain the 
transmission coefficient $\tau$ by relating the oscillation
power at the boundary vertical junction ($n=101$) \emph{with} and
\emph{without} a DB inside the system, as
\begin{equation}
  \label{eq:transmcoeff_sim_def}
  \tau\equiv\frac{P_{\text{ac},101}(\text{with
        DB})}{P_{\text{ac},101}(\text{without DB})}.
\end{equation}

%

\begin{figure}[tb]
  \centering
  \includegraphics[width=6cm]{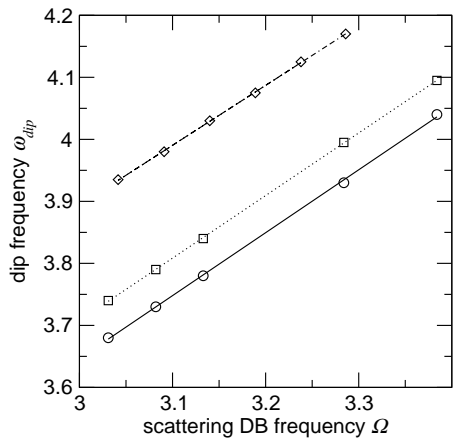}
  \caption{Dip frequency $\omega_\text{dip}$ in the transmission from
    Fig.~\ref{fig:rbtransm_marcussimu}(b) 
    versus DB frequency $\Omega$ for a type A DB
    (two dips, marked by circles and squares) and a type B DB
    (one dip, diamonds). Lines indicate linear fits of the dip frequency to
    Eq.~\eqref{eq:fanodep_numeric}.}
  \label{fig:dipvsbrfreq_nummarcus}
\end{figure}

Fig.~\ref{fig:rbtransm_marcussimu}(b) shows the obtained transmission
for type A scatterer states. Outside the $\omega_3$ band, linear
waves do not propagate, hence any obtained transmission coefficient
is meaningless. Inside the band, we show the transmission through
DBs of three different frequencies. In the first case of small DB frequency
$\Omega_b=2.777$, the transmission curve has a broad
maximum. The DB is practically transparent in the center of the
$\omega_3$ band. If we increase the frequency
of the scattering DB, dips appear in the transmission
at frequencies $\omega_\text{dip}$. 
The values of these frequencies 
change upon variation of the scatterer DB frequency $\Omega_b$.

We systematically varied the scattering DB frequency and computed
various transmission curves. In Fig.~\ref{fig:dipvsbrfreq_nummarcus},
we show the dependence of the positions of the dips in the spectrum
$\omega_\text{dip}$ 
as a function of the scattering DB frequency $\Omega_b$ for type A and
type B states. The obtained dependence is linear and follows
\begin{equation}
  \label{eq:fanodep_numeric}
  \omega_\text{dip}=A_0+A_1\Omega_b.
\end{equation}
For the type A scattering DB, two dips were observed.
For the lower one
we find $A_0=0.601$ and $A_1=1.015$ from a least squares fit, while
the upper resonance satisfies 
$A_0=0.680$ and $A_1=1.009$. In the case of a B type DB, we observe only one
resonance, following  $A_0=0.995$ and
$A_1=0.966$.
 We interpret
$A_0$ as the frequency of a dc local mode, while $A_1$ is practically
equal to unity, and find perfect agreement with Eq.~(\ref{dc}). We
thus conclude that the dips in the transmission spectrum are in fact
Fano resonances.

\begin{figure*}[tb]
  \centering
  \includegraphics[width=1.3\columnwidth]{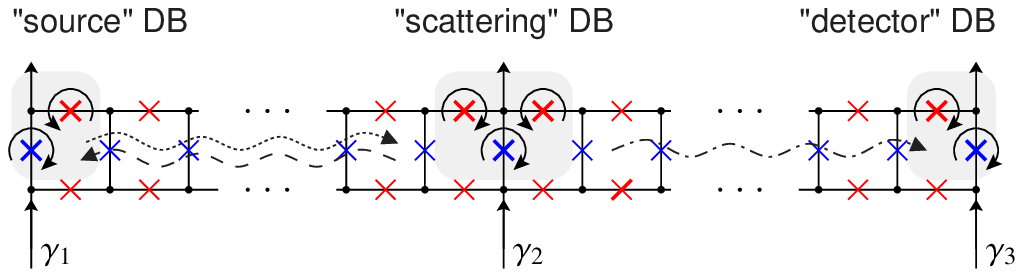}
  \caption{Schematic experimental setup for measuring
  the plasmon scattering by a DB, showing source, scattering, and
  detector DB. The individual DBs are controlled using local bias
  currents $\gamma_1$--$\gamma_3$. Technical details are found in
  Ref.~\cite{msdiss04}.}
  \label{fig7}
\end{figure*}
The persistence of the Fano resonance in the direct simulations of the
finite size Josephson ladder including damping gives a strong
indication for  a possible experimental observation of the phenomenon.
Experimentally, the time-dependent local biasing of one ladder end 
requires microwaves. However, instead of using an external source,
a resonant DB state, as
studied experimentally in Ref.
\cite{schuster01:_obser_josep}, could be used to create linear
waves. A scattering setup should therefore consist of a resonant DB
state used as an emitter (creating linear waves), while another
(non-resonant) DB within a suitable distance would act as a scatterer.
For the detection of radiation we suggest the use of
another set of resistive Josephson junctions inside the ladder, situated at
the far end. These junctions may be seen again as a DB state. 
Their property of quasi-particle detection can be then used
by biasing them close to the superconducting gap voltage
\cite{dayem}. Fig.~\ref{fig7} gives a sketch of the experimental setup
of locally biased DB states used as linear wave source, scatterer, and
detector. Promising experimental studies are under way (see
Ref.~\cite{msdiss04} for details).


\section{Conclusion}
We investigate theoretically the existence of Fano resonances in
wave scattering by DBs in JJLs. Due to the weak 
coupling between open and closed channels of the scattering
potential, generated by a DB, perfect reflection occurs, according
to the theory in the Hamiltonian limit, when multiples of
frequency of DBs match the difference between the frequency of a
dc local mode and some frequency in the spectrum. Numerical
calculations of the transmission coefficient based
on the Floquet approach confirm this analytical
prediction. Moreover, Fano resonances survive
even the presence of dissipation and  external dc bias. 
Direct numerical simulations of the damped and biased system
demonstrate Fano resonances as well.
This is a strong indication that Fano resonances
can be observed experimentally in Josephson junction ladders.

An interesting question is why the direct numerical simulations do not show
a perfect reflection (cf. Fig.~\ref{fig:rbtransm_marcussimu}),
at variance with the computation of the transmission coefficient
based on the Floquet approach (cf. Fig.~\ref{fig6}).
We checked that this is neither an artifact of the numerics, nor 
due to finite size effects. The two methods
(direct simulations and Floquet based approach) differ in
the way boundary conditions are defined. The only explanation we thus
arrive at, is that although the damping is weak, it leads
to a {\it different damping} of the propagating waves in different channels. 
Waves propagating locally around the breather
at different frequencies (i.e. in different channels) are damped
at a different rate, which depends on the frequency of oscillations.
Correspondingly the phase and amplitude relations between the 
partial waves in different channels are changed, leading only
to a partial destructive interference \cite{gurvitz92}.
Alternative explanations involve the broadening of a resonance frequency line
due to damping. The damping causes the transmission to stay
nonzero in the resonance, and changes (broadens) the line width of
the resonance. This also leads to the impossibility of resolving two
closely nearby lying resonances, exactly as we found in our numerical studies
for the type B breather. 

Fano resonances can be considered as a benchmark of dynamical localized
excitations (DBs) in their action on propagating waves in the system.
A similar proposal for the observation of Fano resonances in light-light
scattering in nonlinear optical media 
has been reported in Ref.\cite{gorbach}.
\\
\\
{\bf Acknowledgements}
\\
We thank R. Pinto for a careful reading of the manuscript.
M.V.F. thanks the financial support of SFB 491.

\appendix
\section{Appendix A: Dispersion of linear waves}
Substituting the expressions (\ref{lin_gr}) into system
(\ref{jjl}) together with the transformation (\ref{ttrans}) and
linearizing with respect to the small excitations, we obtain
\begin{eqnarray}
\ddot{\psi}_{n}^v+(\sqrt{1-\gamma^2}-\frac{\alpha^2}{4})\psi_{n}^v&=&\frac{1}{\beta_{L}}(\triangle\psi_{n}^v+\nabla\psi_{n-1}^h-\nabla\tilde{\psi}_{n-1}^h)\nonumber
\\
\ddot{\psi}_{n}^h+(1-\frac{\alpha^2}{4})\psi_{n}^h
&=&-\frac{1}{\eta\beta_{L}}(\nabla\psi_{n}^v+\psi_{n}^h-\tilde{\psi}_{n}^h)
\\
\ddot{\tilde{\psi}}{}_{n}^h+(1-\frac{\alpha^2}{4})\tilde{\psi}^h_n&=&\frac{1}{\eta\beta_{L}}(\nabla\psi_{n}^v+\psi_{n}^h-\tilde{\psi}_{n}^h)\nonumber\;.
\end{eqnarray}
By using the plane wave ansatz
\begin{eqnarray}
\vec{\psi}_n={\mathrm e}^{{\mathrm i}(qn - \omega_qt)}\vec{A}_q\;,
\end{eqnarray}
one obtains that there are three bands. One of them is
dispersionless \beq \omega_1^2(q)=1-\frac{\alpha^2}{4}\;. \eq
For $\alpha=0$ this is the plasma frequency and it is characterized by in-phase
excitations of upper and lower horizontal junctions $A^h=\tilde{A}^h$
and all vertical junctions being at rest $A^v=0$. 
Due to the
absence of dispersion this mode can be excited in each cell with
arbitrary amplitudes, since it does not propagate along the ladder. 

The two
other bands are located above and below it \beq
&&\omega_{2,3}(q)^2=F\mp\sqrt{F^2-G}\;,\nonumber\\
F&=&\frac{1}{2}+\frac{1}{\beta_L\eta}+\frac{1}{2}
\sqrt{1-\gamma^2}+\frac{1}{\beta_L}(1-\cos q)-\frac{\alpha^2}{4}\;,\nonumber\\
G&=&(1+\frac{2}{\beta_L\eta}-\frac{\alpha^2}{4})(\sqrt{1-\gamma^2}-\frac{\alpha^2}{4})+\\
&&+\frac{2}{\beta_L}(1-(1+\frac{\alpha^2}{4})\cos
q-\frac{\alpha^2}{4})\nonumber\;. \eq The dispersion of the second
band $\omega_2(q)$ is weak. 
The polarization vectors of these two branches can be written in a compact form
\beq
 A^h=-\tilde{A}^h\;,\;A^h=
 \frac{{\mathrm e}^{{\mathrm i}q}-1}{\beta_L\eta(1-\omega_{2,3}^2(q)-\alpha^2/4)+2}A^v
\eq

In the Hamiltonian limit $\alpha=0$ and $\gamma=0$ the
band $\omega^2_{2}(q)=1$ becomes dispersionless as well. Its 
polarization vectors are simplified to
out-of-phase excitations of upper and lower
horizontal junctions $A^h=-\tilde{A}_h$ and 
$A^h=\frac{1}{2}(1-{\mathrm e}^{{\mathrm i}q})A^v$. The third
band $\omega_3(q)$ keeps a finite dispersion
 \beq\label{spectr}
\omega^2_3(q)=1+\frac{2}{\beta_L\eta}+\frac{2}{\beta_L}(1-\cos q) \eq
with corresponding in-phase excitations of upper and lower horizontal
junctions and the following relation between vertical and
horizontal ones \beq\label{polarized}
A^h=\tilde{A}^h\;\;,\;\;\;A^v=\eta({\mathrm e}^{-{\mathrm
i}q}-1)A_h\;. \eq

\end{document}